\documentclass[english]{aastex}
\usepackage[T1]{fontenc}
\usepackage{float}
\usepackage{graphicx}
\usepackage{amssymb}

\makeatletter

\providecommand{\tabularnewline}{\\}

\slugcomment{}
\shorttitle{}
\shortauthors{}

\makeatother

\usepackage{babel}

\begin{document}

\title{The Interior Dynamics of Water Planets}

\author{Roger Fu\altaffilmark{1}}

\affil{Earth and Planetary Sciences Dept., Harvard University, 20 Oxford
Street, Cambridge, MA 02138}

\email{rogerfu12@gmail.com}

\author{Richard J. O'Connell}

\affil{Earth and Planetary Sciences Dept., Harvard University, 20 Oxford
Street, Cambridge, MA 02138}

\email{richard\_oconnell@harvard.edu}

\and{}

\author{Dimitar D. Sasselov}

\affil{Harvard-Smithsonian Center for Astrophysics, 60 Garden Street, Cambridge,
MA 02138}

\email{sasselov@cfa.harvard.edu}

\altaffiltext{1}{Corresponding author}
 
\begin{abstract}
The ever-expanding catalog of detected super-Earths calls for theoretical
studies of their properties in the case of a substantial water layer.
This work considers such water planets with a range of masses and
water mass fractions ($2-5\, M_{Earth}$, $0.02-50\%\, H_{2}O$).
First, we model the thermal and dynamical structure of the near-surface
for icy and oceanic surfaces, finding separate regimes where the planet
is expected to maintain a subsurface liquid ocean and where it is
expected to exhibit ice tectonics. Newly discovered exoplanets may
be placed into one of these regimes given estimates of surface temperature,
heat flux, and gravity. Second, we construct a parametrized convection
model for the underlying ice mantle of higher ice phases, finding
that materials released from the silicate - iron core should traverse
the ice mantle on the time scale of $0.1$ to a hundred megayears.
We present the dependence of the overturn times of the ice mantle
and the planetary radius on total mass and water mass fraction. Finally,
we discuss the implications of these internal processes on atmospheric
observables.
\end{abstract}


\section{Introduction}

The discoveries of planetary systems around other stars has now led
us into the realm of planets in the mass range from about 2 to 10-15
$M_{Earth}$, which are expected to be terrestrial in nature and unlike
the gas giants in our Solar System. This new type of planets has been
collectively called \textquotedbl{}super-Earths\textquotedbl{} \citep{ref1}.
Initially, mass is the main physical parameter available. As such,
observers who have discovered nearly a dozen such planets to date
(e.g., \citealt{ref51} and references therein) cannot yet distinguish
between the various interior structures that are expected in this
mass range, i.e. water-rich, dry, or gas-dominated ('mini Neptunes')
types. This is expected to change shortly, with the first super-Earth
discovered in transit, CoRoT-7b \citep{ref2}, and many more anticipated
to come from NASA's Kepler mission, affording derivation of mean densities
and bulk composition. 

Theoretical models of planet formation anticipate large numbers of
water-rich super-Earths (\citealt{ref3,ref4,ref5}). In this paper
we define as water planet any planet in the super-Earth mass range
composed of $>10\%$ $H_{2}O$ by mass around a silicate - metal core
and that lacks a significant gas ($H$ and $He$) layer.

Water planets were proposed by \citet{ref6} and \citet{ref7}, and
modeled with detailed $T$-dependent equations of state (EOSs) by
\citet{ref8}, \citet{ref9}, and \citet{ref10}. Mass-radius relations
have been computed to help distinguish between the different types
of super-Earths when observations become available (e.g. \citealt{ref11},
\citealt{ref12}, \citealt{ref13}, and \citealt{ref14}). \citet{ref15}
and \citet{ref16} considered cold, ice-covered water planets. With
the prospect of high-quality observational data on water planets becoming
available (CoRoT, Kepler, JWST), there is strong motivation to take
the modeling described above to a new level and focus on the geophysical
details of the water layer(s), which is the goal of this work.

This work begins by constructing hydrostatic models that take into
account the effect of ice phase transitions up to Ice X. We model
nine ocean planets with total mass ranging from $2$ to $10\, M_{Earth}$
and water mass percents from $0.02\%$ (Earth-like) to $50\%$. Section
\ref{sec:Hydostatic-Models} explains the creation of these models. 

Section \ref{sec:Thermal-Models} presents a pair of models to constrain
the thermal and dynamical structure within the $H_{2}O$ layer of
these planets. We consider a variety of imposed conditions such as
surface temperature (above and below freezing) and heat flux. The
first model governs the behavior of the near-surface region and considers
both conduction and solid-state convection as modes of heat transport.
The second is a parameterized model for the underlying mantle of higher-phase
ices. The results from these models are presented in Section \ref{sec:Results}. 

We follow with a discussion (Section \ref{sec:Discussion}) of remaining
uncertainties and of the observable implications of these thermal
models, especially regarding the transport of material from the silicate
- metal core to the surface.

\section{Hydrostatic Models\label{sec:Hydostatic-Models}}

We construct hydrostatic models of nine hypothetical planets with
total masses of 2, 5, and 10 $M_{Earth}$. For each mass value, we
consider a dry, Earth-like (simplified to be 32\% $Fe$, 68\% $MgSiO_{2}$,
and 0.02\% free $H_{2}O$) case and two cases with $25\%$ and $50\%$
water by mass around a core with Earth-like composition. These planets
are simplified to contain up to five distinct compositional layers:
iron, $MgSiO_{3}$ perovskite, Ice X, Ice VII, and Ice Ih. 

We use the canonical equations for a spherical body in hydrostatic
equilibrium in conjunction with an EOS for each material to obtain
the preliminary dimensions of the exoplanets. We follow the results
of \citet{ref13} and choose the Vinet and fourth-order Birch-Murnaghan
EOS (\citealt{ref17,ref18}) for iron and $MgSiO_{3}$ perovskite,
respectively. We choose the third-order Birch-Murnaghan EOS for all
three water ice phases. Our choices of EOS and parameters used are
summarized in Table \ref{tab:parameters for statics}. The effect
of the choice of EOS is minor for our purpose here (e.g., \citealt{ref8,ref13,ref10}).
Following these studies, we use isothermal EOSs for all regions.

Several approximations are employed here. Clearly, the pure chemical
composition assumed in each of the three regions is non-physical,
but its associated error is acceptable. For instance, the replacement
of 25\% of the Mg by Fe in the post-perovskite $MgSiO_{3}$ structure
results in a 2.5\% increase in density \citep{ref52}. Any changes
in the property of the silicate mantle due to different internal water
content compared to that of the Earth are neglected. 

As for phase changes, we neglect any transitions in the Fe core. Due
to the lack of necessary parameters for higher phases, the entirety
of the silicate mantle is assumed to be in the perovskite phase of
$MgSiO_{3}$. The higher-pressure post-perovskite phase is stable
from around 125 GPa up to a currently unknown phase transition (e.g.
\citealt{ref25}) or to its dissociation into $MgO$ and $SiO_{2}$
at about 1000 GPa \citep{ref27}, making it the predominant component
of the silicate layer of all of our modeled planets except the $2\, M_{Earth}$,
Earth-like one. Even so, we find that this as well as the above described
approximations are acceptable, given that our focus in this work is
on the overlying water layers. 

Following the experimental phase diagram for high pressure ice phases
(\citealt{ref28,ref29,ref30}), we use the elastic parameters of Ice
VII between $2.2$ and $60\, GPa$, while Ice X is assumed to exist
above $60\, GPa$. In fact, no static laboratory experiments constrain
the behavior of ice at pressures above $170\, GPa$, while the pressure
at the silicate-ice boundary in the 5 and 10 $M_{Earth}$, 50\% water
planets reaches $212$ and $437\, GPa$, respectively. Recent experiments
by \citet{ref29} have led to the possibility of a higher ice phase
at $150\, GPa$ and above. However, for lack of further experimental
data, we approximate all potential higher phase ices as Ice X. 

In our hydrostatic models, we ignore the possible existence (depending
on thermal profile) of other phases of water (e.g. liquid, Ice III,
Ice VI) between the Ice Ih and Ice VII regions. Due to the relatively
small thickness of these near-surface water layers, the effect of
these approximations on key parameters such as surface gravity $g_{s}$
and planetary radius $r_{p}$ is negligible. 

\begin{figure}[H]
\includegraphics[width=0.9\textwidth]{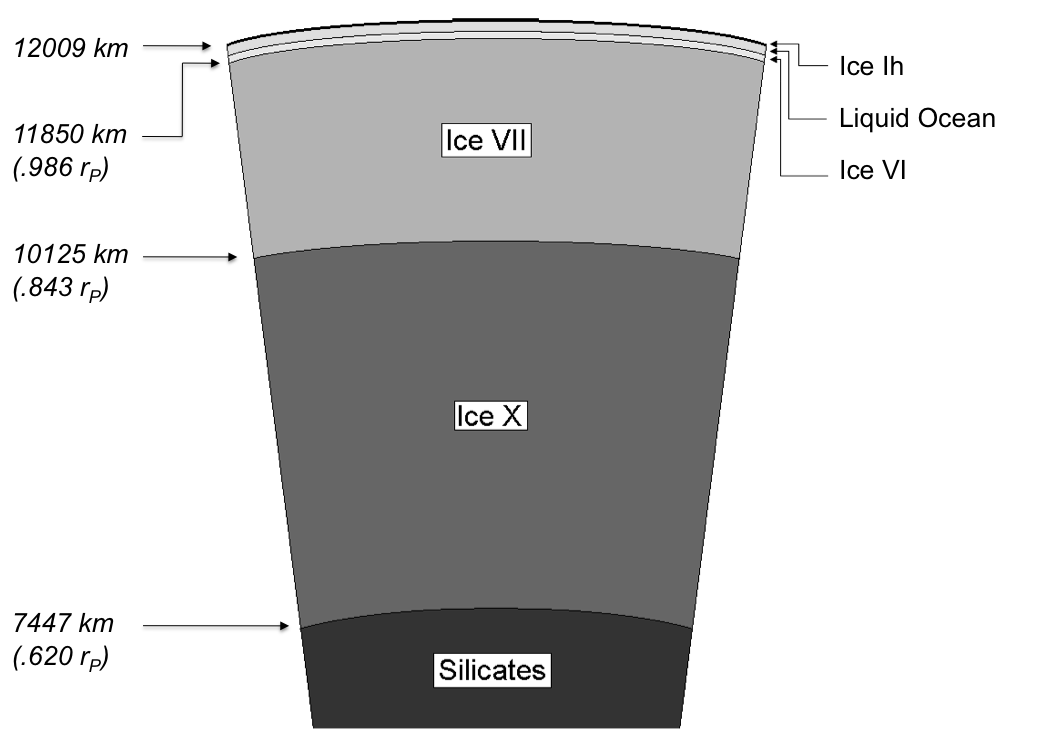}

\caption{The $H_{2}O$ region of the $5\, M_{Earth}$, $50\%\, H_{2}O$ planet
drawn to scale. This configuration estimates heat flux according to
Equation \ref{eq:q_s scaling law}.\label{fig:scale drawing}}

\end{figure}

A scaled diagram of the $H_{2}O$ region of our $5\, M_{Earth}$,
$50\%\, H_{2}O$ planet is shown in Figure \ref{fig:scale drawing}.
Note that the details in the Ice VI, liquid water, and Ice Ih layers
were found after the thermal modeling in the subsequent sections.

\section{Thermal Models\label{sec:Thermal-Models}}

We include two sets of thermal models. The first models the near-surface
region of the planet, both in the case of a frozen surface and a liquid
surface. The second models convection in the underlying mantle of
higher-phase ices.

\subsection{Near surface convection}

We consider the case where the surface temperature is below the freezing
point. We model both conductive and convective heat flow in the near-surface
region. We define the superficial ice layer where heat transport is
solely due to conduction as the {}``crust.'' For the thickness of
the crust, we use the simple expression for a conductive thermal profile:

\begin{equation}
D=\frac{k(T_{B}-T_{S})}{q_{S}}\label{eq:conductive crust thickness}\end{equation}

Where $D$ is the thickness of the crust, $T_{B}$ and $T_{S}$ are
the temperatures at the base and at the surface, $q_{S}$ is the surface
heat flux, and $k$ is the thermal conductivity of Ice Ih, equal to
$3.3\, Wm^{-1}K^{-1}$ \citep{ref31}. $T_{S}$ and $q_{S}$ are left
as the independent variables in this analysis. The former is determined
by the planet's insolation, albedo, and atmospheric effects, while
heat flux can be estimated by scaling the planetary heat output with
the total mass of silicates and metals ($M_{eff}$) compared to the
Earth, and then adjusting this quantity to the desired radius ($r$):

\begin{equation}
q_{s}=q_{Earth}\frac{M_{eff}}{M_{Earth}}(\frac{r}{r_{Earth}})^{-2}\label{eq:q_s scaling law}\end{equation}

Where $q_{Earth}=0.087\, W/m^{2}$ \citep{ref71}. This estimate for
heat flux is only used in this work for analyzing the convection of
the ice mantle (Section \ref{sub:The-ice-mantle}). 

We evaluate the convective stability of the surface ice layer by finding
its Rayleigh number ($Ra$) and comparing it to a critical Rayleigh
number ($Ra_{crit}$). The former is given by:

\begin{equation}
Ra=\frac{g\alpha\rho D^{3}(T_{B}-T_{S})}{\kappa\eta(T_{B},T_{S})}\label{eq:rayleigh}\end{equation}

Where $g$ is the surface gravity, $\alpha=1.6\times10^{-4}\, K^{-1}$
is the thermal expansivity of Ice Ih, $\rho=917\, kg\, m^{-3}$ is
its density, and $\kappa=3.7\times10^{-6}\, m^{2}s^{-1}$ is its thermal
diffusivity \citep{ref31}. The viscosity $\eta$ is strongly temperature
dependent. Given the low pressure regime found in the crust, we assume
a Newtonian, volume diffusion-dominated rheology following the work
of \citet{ref33}:

\begin{equation}
\eta(T)=\eta_{0}\exp[\frac{Q^{*}}{RT_{\eta}}]\label{eq:viscosity}\end{equation}

We adopt values of $139$ $Pa\cdot s$ and $56.7$ $kJ/mol$ for $\eta_{0}$
and $Q^{*}$, respectively, corresponding to a grain size of $\approx0.2\, mm$
for our temperature range \citep{ref53}. 

We note that the stability of the conductive boundary layer has been
used to derive the relation $Nu=A\cdot Ra^{\beta}$, where $Nu$ is
the Nusselt number and $A$ and $\beta$ are constants. Scaling relations
of this form are used to relate the convective heat flux to the parameters
governing convection \citep{ref74}, and they yield the same result
as more complete boundary layer models (\citealp{ref75}; \citealp{ref71};
\citealp{ref72}). Where applicable (i.e. in the case of convection
in an Ice Ih shell overlying a liquid ocean), this scaling has been
verified to corroborate our results. 

We evaluate Equations \ref{eq:conductive crust thickness}, \ref{eq:rayleigh}
and \ref{eq:viscosity} under two distinct formulations. For relatively
high $T_{S}$ ($\gtrsim190\,-\,220\, K$, depending on curve being
calculated), we find that the viscosity contrast is relatively small
($\mbox{\ensuremath{\eta}}_{max}/\eta_{min}\lesssim10^{3}$). We therefore
employ the small viscosity contrast (SVC) presciption, with characteristic
viscosity taken to be the value corresponding to the average temperature:
$T_{\eta,\, SVC}=(T_{B}+T_{S})/2$. This approximation has been found
to be appropriate for viscosity contrasts of up to $\eta_{max}/\eta_{min}\approx10^{4}$
\citep{ref57}. In this regime, we use a critical Rayleigh number
of $Ra_{crit,\, SVC}=2000$, corresponding to viscosity contrasts
of $10^{3}$ \citep{ref36}.

For relatively low $T_{S}$ and correspondingly high viscosity contrasts,
we use the stagnant-lid formulation to describe convection. In this
case, convection is limited to an adiabatic zone in the lowest part
of the ice layer. The characteristic viscosity of the convection cell
is found by evaluating Equation \ref{eq:viscosity} with the temperature
at the bottom of the conductive crust: $T_{\eta,\, SL}=T_{B}$ \citep{ref54}.
The corresponding critical Rayleigh number is dependent on the magnitude
of the viscosity contrast: $Ra_{crit,\, SL}=20.9p^{4}$ where $p$
reflects the logarithm of the viscosity contrast between the top and
bottom of the full ice layer \citep{ref56}:

\begin{equation}
p=\frac{Q^{*}}{R}(\frac{T_{B}-T_{S}}{T_{i}^{2}})\end{equation}

Where $T_{i}$ is the temperature in the actively convecting region.
It is cooler than $T_{B}$ by a small amount on the order of $10\, K$:
($T_{B}-T_{i}\approx RT_{i}^{2}/Q^{*}$).

Using the appropriate formulation for the given viscosity contrast,
we solve for the maximum thickness of the conductive crust and the
bottom temperature $T_{B}$ as functions of the imposed parameters
$q_{S}$, $T_{S}$, and $g$. If this maximum conductive thickness
is greater than that of the Ice Ih layer before it meets the melting
curve, then our crust directly overlies a liquid ocean, which has
an adiabatic temperature profile. Otherwise, we find solid-state convection
immediately below the crust.

In the case of a liquid surface, we model the temperature gradient
in the surface ocean to be adiabatic. The thickness of this ocean
is found by identifying the intersection of this adiabat with the
freezing curve of water at depth.

\subsection{The ice mantle \label{sub:The-ice-mantle}}

\begin{figure}[H]
\includegraphics[width=0.9\textwidth]{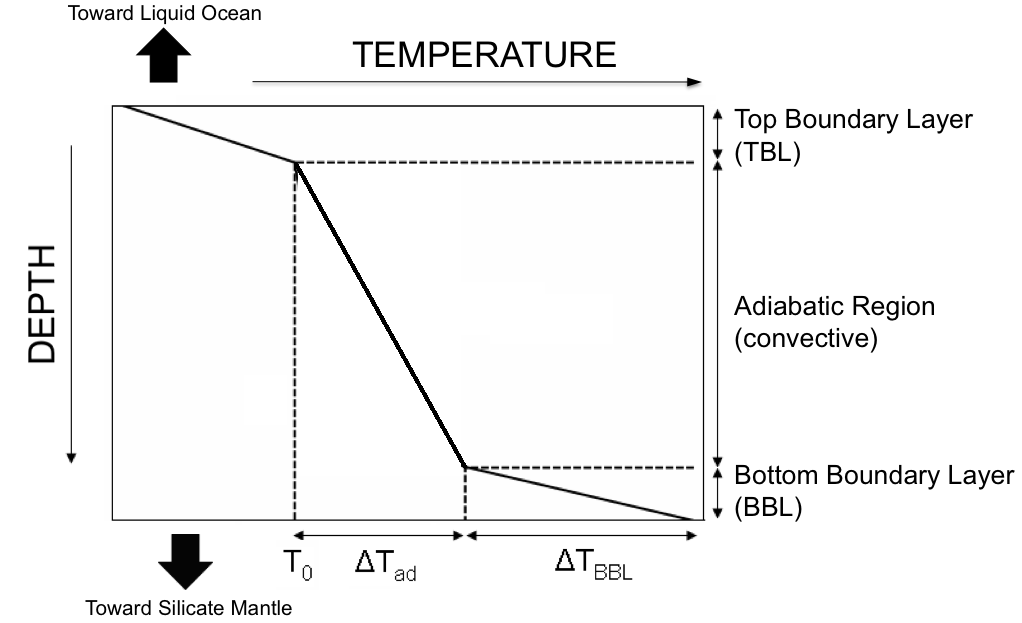}

\caption{The qualitative features of the thermal profile within the ice mantle.
The temperature in the TBL is bound to the melting curve while that
of the BBL is conductive.\label{fig:The-assumed-shape}}

\end{figure}

Regardless of the near-surface structure, convection is always present
in the underlying layers of higher phase ices (henceforth called the
{}``ice mantle''). We assume for now and show below that convection
within the ice mantle is not partitioned by phase transitions. We
also assume a rectangular geometry for the convective layer, a valid
approximation given that the analysis focuses on only the boundary
layers. For the case where the ice mantle underlies a liquid ocean,
we adopt a thermal profile as shown in Figure \ref{fig:The-assumed-shape}.
This thermal profile is partitioned into an adiabatic region between
two boundary layers. The Bottom Boundary Layer (BBL) is a conductive
zone that corresponds to the thin region where the cold material recently
arrived from above is slowly heated by conduction from the underlying
silicates until it rises again. 

On the other hand, the temperature profile in the top boundary layer
(TBL) is constrained to follow the melt curve. The dynamical characteristic
of this region is outlined further in the discussion (Section 5).
The top interface temperature $T_{T}$ and depth $z_{T}$ correspond
to the temperature and location of the bottom of the liquid ocean
and are constrained by the above considerations of the near-surface
thermal profile. To constrain the quantities $T_{0}$, $\Delta T_{ad}$,
and $\Delta T_{BBL}$ (see Figure \ref{fig:The-assumed-shape}), we
construct a convection model that uses three independent relations:
one governing each boundary layer and one describing the adiabat.
We note that we do not explicitly use any existing Nusselt - Rayleigh
number scaling laws. However, our model considers precisely the same
physics that forms the basis of such scaling laws- namely by assuming
that both boundary layers are at the verge of convective stability
(see \citealp{ref72}, the Appendix for a detailed derivation). 

First, we examine the convective stability of the TBL. In all six
of our examined ice mantles (of the $25\%$ and $50\%$ $H_{2}O$
planets), when approximating heat flux by Equation \ref{eq:q_s scaling law},
the TBL is composed entirely of Ice VI. The thickness ($\delta_{TBL}$)
and bottom temperature ($T_{0}$) in the TBL is therefore related
simply by following the melt curve of Ice VI:

\begin{equation}
\delta_{TBL}=\frac{T_{0}-T_{VI}}{m_{VI}}-z_{T}\label{eq:delta TBL}\end{equation}

Where $T_{VI}$ and $m_{VI}$ are the $T$-intercept and slope of
the Ice VI melt curve, respectively, and $z_{T}$ is the depth of
the bottom of the liquid ocean. The Rayleigh number of the TBL can
then be calculated and set equal to the critical Rayleigh Number ($Ra_{crit}$),
chosen to be $1000$ to reflect the mild viscosity contrast within
the layer \citep{ref71}:

\begin{equation}
Ra_{TBL}=Ra_{crit}=\frac{g\alpha\rho(T_{0}-T_{T})\delta_{TBL}^{3}}{\kappa\eta_{TBL}}\label{eq:Ra TBL}\end{equation}

Where $\eta_{TBL}$ is the viscosity in TBL evaluated for the average
temperature and pressure, as appropriate for smaller viscosity contrast
\citep{ref35}. $T_{0}$ may be obtained from Equations \ref{eq:delta TBL}
and \ref{eq:Ra TBL}, allowing for the evaluation of the adiabatic
temperature gradient via the following relationship \citep{ref71}:

\begin{equation}
\frac{dT(z)}{dz}\approx\frac{T(z)\, g\alpha(P(z),\, T(z))}{C_{p}(T(z))}\label{eq:new adiabat}\end{equation}

Where $C_{P}(T)$ is the isobaric specific heat capacity. Equation
\ref{eq:new adiabat} can be integrated numerically from bottom of
the TBL to the top of the BBL to find the temperature change in the
adiabatic regime $\Delta T_{ad}$ as a function of $T_{0}$. 

Finally, we evaluate the Rayleigh number of the BBL and set it to
the critical Rayleigh number $Ra_{crit}$, again chosen to be $1000$.
Defining $\delta_{BBL}$ to be the thickness of the BBL, we can write
the Rayleigh number and the heat flux as:

\begin{equation}
Ra_{BBL}=Ra_{crit}=\frac{g\alpha\rho\Delta T_{BBL}\delta_{BBL}^{3}}{\kappa\eta_{BBL}}\label{eq:BBL 1}\end{equation}

\begin{equation}
\delta_{BBL}=\frac{k\Delta T_{BBL}}{q_{BBL}}\label{eq:BBL 2}\end{equation}

Where $\eta_{BBL}$ is the BBL viscosity, which is dependent on the
value of $T_{0}$ and $T_{ad}$, and is once again calculated for
small internal viscosity contrast. Combining Equations \ref{eq:BBL 1}
and \ref{eq:BBL 2} allows the calculation of the final unknown, $\Delta T_{BBL}$. 

Due to the much higher pressure and temperatures in the ice mantle,
the assumptions of Newtonian rheology and constant $\alpha$ as in
our near-surface analysis are insufficient. Under the stress regime
in our ice mantles (several $MPa$, see below), dislocation creep,
which has a significant stress dependence, is expected to dominate
\citep{ref73}. Furthermore, the viscosity change between known ice
phases under this stress regime is not great ($<10^{2}$). Hence,
we use a dislocation creep model for the viscosity of Ice VI, the
highest pressure phase measured to date, for the ice mantle \citep{ref38}:

\begin{equation}
\eta(P,\, T)=6.65\times10^{19}\,\sigma^{-3.5}\,\exp[-(E^{*}+PV^{*})/RT]\label{eq:VI viscosity}\end{equation}

Where $\sigma$ is a characteristic shear stress, $R$ is the ideal
gas constant, and $E^{*}$ and $V^{*}$ are the activation energy
and volume, measured to be $110\, kJ/mol$ and $1.1\times10^{-5}\, m^{3}/mol$,
respectively. Owing to the lower viscosity of ice compared to silicates,
the shear stress is approximated to be a value somewhat lower than
that of the Earth's mantle, $2\times10^{6}\, Pa$. This value is self-consistent
with our results. The activation volume $V^{*}$ itself changes with
depth; it can be approximated to shrink with pressure as a vacancy
in the material. Following the analysis of \citet{ref39}, the effective
bulk modulus of this vacancy is:

\begin{equation}
K^{*}=V^{*}\frac{\partial P}{\partial V^{*}}=\frac{2(1-2\nu)}{3(1-\nu)}K\label{eq:vstar calc}\end{equation}

Where $K$ is the bulk modulus of the material and $\nu=0.35$ is
the Poisson ratio. By expanding $K$ to first order in pressure and
integrating Equation \ref{eq:vstar calc}, we obtain the activation
volume as a function of pressure:

\begin{equation}
V^{*}(P)=V_{0}^{*}(\frac{K_{0}^{*}+K'^{*}P}{K_{0}^{*}+K'^{*}P_{0}})^{-\frac{1}{K'^{*}}}\label{eq:Vstar of P}\end{equation}

Combining Equations \ref{eq:VI viscosity} and \ref{eq:Vstar of P}
gives viscosity as a function of depth for a specified thermal profile.
Under this formulation for viscosity, our results show that maximum
viscosity contrasts in the TBL are on the order of $10^{2}$, while
that of the BBL are similar in most cases, except for the $2\, M_{Earth}$
planets, where the contrasts are between $10^{5}$ and $10^{7}$.
Despite this, we maintain the use of the small viscosity contrast
formulation for BBL viscosity due to (1) the fact that varying $ $$\eta_{BBL}$
through its entire possible range does not change our ultimate results
(overturn timescale, etc) by more than a factor of a few and (2) the
lack of existing theory for bottom boundary layers with high viscosity
contrast. Note that the viscosity contrasts within the TBL are much
smaller than values for the icy crust due to the viscosity-reducing
effects of pressure and the shallower thermal gradient due to the
melting curve. 

As for the depth-dependent parameters $\alpha(P,\, T)$ and $C_{P}(T)$,
we use results obtained for Ice VII. \citet{ref40} have produced
the following expression for $\alpha(P,\, T)$ , in units of $K^{-1}$:

\begin{equation}
\alpha(P,\, T)=(3.9\times10^{-4}+1.5\times10^{-6}T)(1+\frac{K'_{0T}}{K_{0T}}P)^{-0.9}\label{eq: thermal expansion for ice VII}\end{equation}

We again use the results from \citet{ref40} to obtain $C_{P}$ in
units of $J\, kg^{-1}\, K^{-1}$:

\begin{equation}
C_{p}(P,\, T)=4027.22+0.168T-8.011\times10^{7}\frac{1}{T^{2}}\end{equation}

Finally, we test throughgoing convection at phase transitions throughout
the ice mantle. A sufficiently endothermic phase transition (i.e.
the Clapeyron slope, $\gamma$, is sufficiently negative) can partition
convection by introducing temperature-driven topography to the phase
boundary. We evaluate the dimensionless {}``phase buoyancy parameter''
($P_{b}$) for all endothermic transitions, given by \citep{ref41}:

\begin{equation}
P_{b}\equiv\bar{\gamma}\frac{Rb}{Ra}=\gamma\frac{\Delta\rho}{\bar{\rho}^{2}gh\alpha}\label{eq:P}\end{equation}

Where $Rb$ is the boundary Rayleigh number which uses a phase-driven
instead of thermal-driven density contrast, $\gamma$ is the Clapeyron
slope, $\Delta\rho$ is the density contrast across the transition,
$\bar{\rho}$ is the mean density, and $h$ is the full thickness
of the unpartitioned convective layer. Our values for $\Delta\rho$
are calculated from EOSs for the relevant ice phases, while $\gamma$
was obtained directly from the literature or calculated with published
triple point data. See Table \ref{tab:The-buoyancy-parameter} for
a listing of relevant references.

\section{Results\label{sec:Results}}

We obtain mass-radius relationships for water planets from our hydrostatic
models. Following the work of \citet{ref11}, we fit our results to
a function of total mass of the form:

\begin{equation}
\frac{r_{P}}{r_{Earth}}=A(\frac{M_{P}}{M_{Earth}})^{B}\label{eq:scaling mr}\end{equation}

With values for $A$ and $B$ given for a range of water compositions
in Table \ref{tab:Constants-for-the}. The silicate - metal ratio
in each case is $2.09$. We find that Ice X lies at the bottom of
the $H_{2}O$ shell of the $25\%$ and $50\%$ $H_{2}O$ planets except
for the $2\, M_{Earth}$, $25\%\, H_{2}O$ case (see Table \ref{tab:Pressures-at-the}).
In the cases of Earth-like water mass fraction, we find either Ice
Ih or liquid water at the $H_{2}O$ - silicate boundary, depending
on the near-surface temperature profile (see below).

For the planets with ice phase transitions ($25\%$ and $50\%\, H_{2}O$
planets), our analysis of convective breakthrough revealed no phase
transitions that should partition convection. As an example, the relevant
parameters and $P_{b}$ values for the $5\, M_{Earth}$, $50\%\, H_{2}O$
exoplanet ($h\approx4400\, km$) are listed in Table \ref{tab:The-buoyancy-parameter}.
Note that our predictions of convective breakthrough at the Ice Ih
- III and Ice II - V boundaries diverge from the corresponding results
for the icy satellites \citep{ref35}. This is due to the much greater
thickness $h$ of the convective cell in our exoplanet. For a much
smaller exoplanet closer to the Galilean satellite instead of super-Earth
regime, given an ice mantle thickness on the order of a few hundred
$km$, these two transitions may hinder convection.

\begin{figure}[H]
\includegraphics[width=0.8\textwidth]{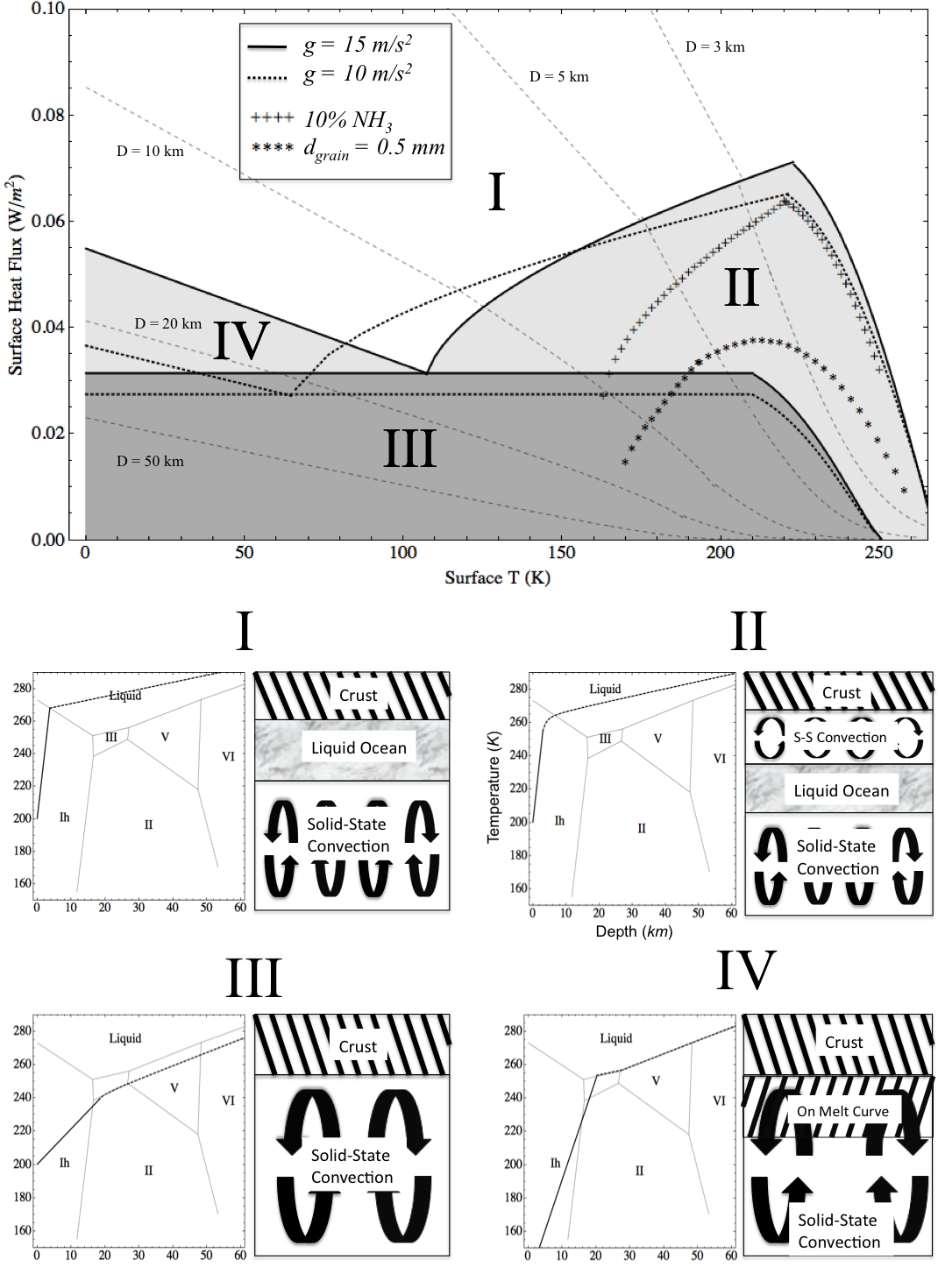}

\caption{{\footnotesize The four Regimes of the crust as a function of $T_{S}$
and $q_{S}$ for the $g=15\, m/s^{2}$ and $10\, m/s^{2}$ cases.
Contours of constant crustal thickness for $g=15\, m/s^{2}$ are plotted
as light dashed lines in the top diagram. See Section \ref{sec:Results}
for explanation of the boundaries of each regime. Depth scale in the
phase diagrams are based on the $5M_{Earth}$, $50\%\, H_{2}O$ planet.
The cartoon cross-sections are not drawn to scale. In the liquid surface
case, the cross-section would appear as in Regime I, only without
the {}``Crust'' section; its thermal profile would begin at an above-freezing
temperature at the surface and be adiabatic throughout the liquid
ocean. \label{fig:The-four-regimes}}}

\end{figure}

Our models of the frozen surface case reveal that four qualitatively
distinct dynamical Regimes exist for the near-surface region. For
a given rheology, the Regimes are determined by the surface temperature
$T_{S}$, heat flux $q_{S}$, and, to a lesser extent, gravity $g$.
The parameter space in which one finds each of the four layering regimes
is shown for the $g=15\, m/s^{2}$ and $g=10\, m/s^{2}$ cases in
Figure \ref{fig:The-four-regimes}. A liquid ocean is present for
Regimes I and II. In Regime I, the Ice Ih layer overlies a liquid
water ocean and is not of sufficient thickness to convect. In Regime
II, the temperature profile still intersects the melting curve within
the Ice Ih region, thus maintaining the presence of an ocean. However,
the Ice Ih region in this case is convectively unstable, and some
solid-state convection is found at the base of this region. The boundary
between Regime I and Regime II is calculated by setting the Rayleigh
number of the strictly Ice Ih region above the ocean layer to the
critical Rayleigh number ($Ra_{crit}$): if the Rayleigh number is
above $Ra_{crit}$, the planet is in Regime II. The kink in this curve
at $\approx225\, K$ is due to a transition between the usage of stagnant-lid
formulation for lower $T_{S}$ and small viscosity contrast formulation
for higher values. At lower $T_{S}$ values, increasing the temperature
lowers the viscosity and expands the parameter space where convection
occurs. However, as $T_{S}$ continues to increase, eventually the
thinning of the crust inhibits convection, hence the characteristic
{}``humped'' shape of the Regime I - II boundary. 

In Regime III, the {}``turn off point'' in the temperature profile-
the point at which the switch is made from conductive to convective
heat transfer- is below the minimum temperature of the melting curve
($251\, K$). Therefore, we find no liquid ocean. Instead, given the
penetration of the Ice Ih - III and Ice Ih - II interfaces, we find
a convection cell immediately below the crust that extends down to
the silicate - ice boundary. Finally, in Regime IV, a thick, convectively
stable crust extends below the region occupied by Ice Ih and meets
the melting curve before the onset of convection. The temperature
profile, when it reaches the melting temperature, is bounded by the
melting curve. This must be the case: the temperature cannot be higher
than the melting curve since the liquid would rapidly carry away heat,
refreezing the material. The temperature profile must therefore follow
the melting curve until this transition layer becomes convectively
unstable, at which point it dives to greater depths along an adiabat.
The dynamical properties of this region bound to the melt curve are
considered further below under Discussion. 

The curve separating Regime III from Regimes II and IV is the contour
for a constant {}``turn off point'' temperature of $251\, K$. Meanwhile,
the line separating Regimes I and IV corresponds to the scenario where
the conductive portion of the temperature profile intersects the melt
curve at the Liquid - Ice Ih - Ice III triple point: $T=251\, K$
and $P=210\, MPa$. If the intersection with the melt curve occurs
at lower pressure (above the Regime I - IV boundary in Figure \ref{fig:The-four-regimes}),
the temperature profile intersects the melting curve within the Ice
Ih domain and therefore falls into Regime I.

Our formulation for Newtonian viscosity (Equation \ref{eq:viscosity})
is subject to uncertainty due to the unknown characteristic grain
size in the Ice Ih crust of exoplanets. Compared to the $\approx0.2\, mm$
we used in our analysis, Galileo observations of Europa has produced
estimates of $0.1\, mm$ for surficial grain size and estimates of
several millimeters in the underlying regions \citep{ref59}. Assuming
that volume diffusion continues to be the dominant creep mechanism,
increasing the grain size to $0.5\, mm$ increases the constant $\eta_{0}$
in Equation \ref{eq:viscosity} by a factor of six ($\eta\propto d^{2}$)
and lowers the required $q_{S}$ value for all Regime boundaries by
about a factor of two, although their $T_{S}$ coordinate remains
essentially identical (see Figure \ref{fig:The-four-regimes}).  The
uncertainty in grain size, and hence Ice Ih viscosity, is the single
greatest uncertainty in our analysis of the ice crust, and care should
be taken to consider a range of possible grain sizes when attempting
to characterize the surface of a specific water planet. 

A further source of uncertainty lies in the possibility of the presence
of solutes or other volatiles in significant quantities within the
ocean layer, thereby depressing the freezing point of the water-solute
mixture, resulting in thinner and colder crusts, possibly hindering
convection. We test the case of a water-ammonia mixture by adopting
the phase diagram of the $H_{2}O$ - $NH_{3}$ system as proposed
by \citet{ref70}. Assuming $NH_{3}$ concentration of $10\%$, we
find a modest correction to the Regime I - II boundary curve (see
Figure \ref{fig:The-four-regimes}).

As pointed out by the referee, we ignore the transition regime between
the low viscosity contrast and stagnant-lid convection regimes. Given
that our calculations yield Rayleigh numbers on the order of $10^{5}$
to $10^{6}$ for the Ice Ih shell in Regime II and $>10^{6}$ for
Regimes III and IV (equal to $Ra$ of the full ice mantle convection
in these latter cases), the surface temperature range over which the
transitional regime is valid is between $T_{S}\approx190\, K$ and
$235\, K$. As the plausible deviation from our interpolation inside
this narrow range is small compared to the magnitude of the other
uncertainties that exist, we accept our current results as sufficiently
accurate while noting that fuller consideration of the transitional
regime may be worthwhile in a future work. 

Although it is not plotted in Figure \ref{fig:The-four-regimes},
we imagine a separate regime where $T_{S}$ is above freezing, resulting
in a liquid ocean at the surface. The temperature profile in such
an ocean is adiabatic until it freezes into higher phase ices at a
depth on the order of $10$s to a few $100$s of $km$. 

\begin{figure}[H]
\includegraphics[width=0.9\textwidth]{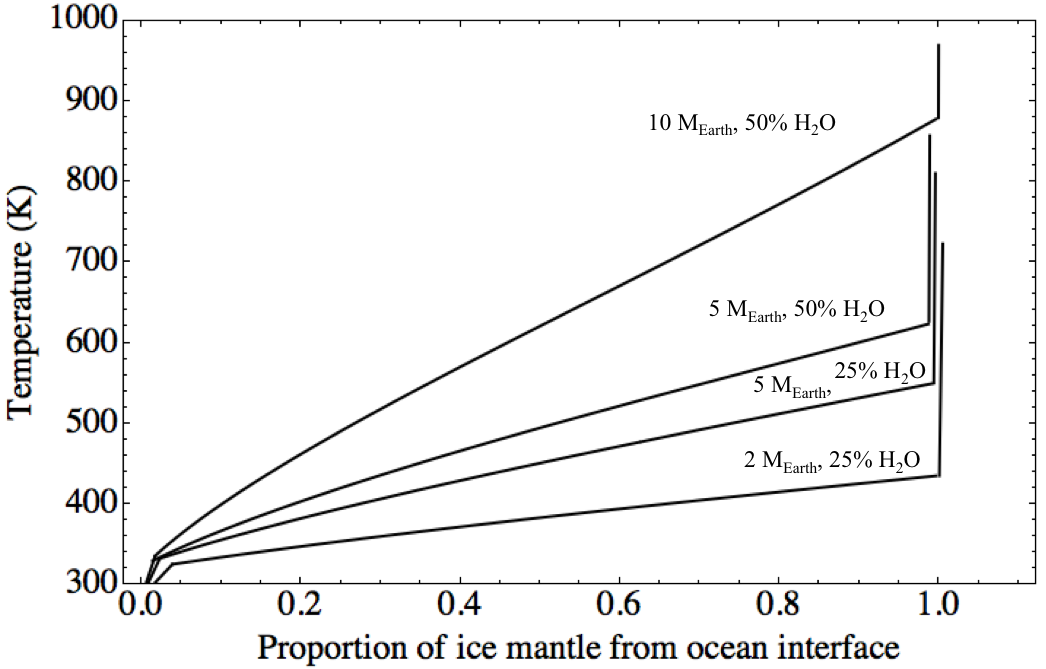}

\caption{Complete temperature profile of the ice mantle of four selected water
planets plotted against normalized depth from the ocean - Ice VI contact.
True mantle thicknesses are: $1910\, km$, $2460\, km$, $4500\, km$,
and $5425\, km$ for the four planets plotted in ascending order as
seen in the figure (i.e. $1910\, km$ corresponds to the $2\, M_{Earth}$,
$25\%\, H_{2}O$ planet, etc). The locations of the bottom boundary
layers have been staggered in order to show the magnitude of temperature
change. \label{fig:Complete-ice-mantle}}

\end{figure}

We apply our convection model for the ice mantle underlying a liquid
ocean to the $25\%$ and $50\%\, H_{2}O$ exoplanets with $T_{S}=250\, K$.
The near surface structure is first determined in each case, yielding
the result the crust in all these cases fall into Regime I. As an
example, in the $5\, M_{Earth}$, $50\%\, H_{2}O$ case we find that
an ocean of $62\, km$ thickness underlies a brittle crust only $1.2\, km$
thick. The ocean bottom temperature is $T_{T}=283\, K$; this figure
is virtually constant for all six cases to within two degrees. The
full solution to the temperature profile in the ice mantle for selected
exoplanets is shown in Figure \ref{fig:Complete-ice-mantle}. We find
that due to the smaller adiabatic temperature change and the correspondingly
lower temperature at the top of the BBL, $\Delta T_{BBL}$ is significantly
larger for planets with the smallest $H_{2}O$ shells. 

We can estimate the full ice mantle Rayleigh number of each planet.
As described further in the Discussion section, finding a characteristic
viscosity for the ice mantle is elusive due to its complex depth-dependence,
reaching a viscosity maximum in mid-layer. We employ a simple approximation
by taking the logarithmic average of the the viscosity throughout
the ice mantle. This estimate results in Rayleigh numbers for the
ice mantle between $10^{5}$ and $10^{9}$, with the smallest Rayleigh
numbers corresponding to the for the least massive planets with the
thinnest water shell. 

The overturn time of the ice mantle (i.e. time needed for material
from the silicate - ice boundary to reach the top of the ice mantle)
can now be calculated. \citet{ref71} have found by theoretical considerations
that, given a cell aspect ratio of unity, the maximum velocity of
convective material should depend on the full mantle Rayleigh number
$Ra_{m}$:

\begin{equation}
u_{0}=0.271\frac{\kappa}{h}Ra_{m}^{2/3}\label{eq:traversal v}\end{equation}

The overturn time scale is then simply calculated by dividing the
thickness of the ice mantle by $u_{0}$. The results of this calculation
for a range of planetary masses and water mass fractions, as well
as their uncertainties, are shown in Table \ref{tab:Overturn-table}.
A word about the associated errors is found in Discussion.

\section{Discussion\label{sec:Discussion}}

If a water planet has a frozen surface, the Regime of the crust has
direct impact on observable features. In Regimes I and II, the presence
of a liquid ocean layer is expected to bias the set of outgassed compounds
towards species that are water-transportable, since, although solid-state
convection through the mantle of higher ice phases can dredge up material
expelled from the silicate - metal core, predominantly materials that
can be dissolved or suspended in liquid water are able to traverse
the ocean layer. A more detailed, quantitative analysis of this selection
process is left for future work. 

In Regimes II, III, and IV, the presence of solid-state convection
immediately below the brittle crust may lead to the existence of ice
tectonics. This provides a means by which the chemistry of the convection
cells can continuously gain a window to the surface, altering the
atmospheric observables of the exoplanet \citep{ref46}. A simple
test for the presence of ice tectonics is to compare the tensile strength
of Ice Ih as found on the surfaces of icy satellites ($\approx1\, MPa$,
\citealt{ref47}) and the expected tensile stress imposed on the crust
by the convection cell. The approximate stress imposed by underlying
convection is \citep{ref35}: 

\begin{equation}
\sigma_{conv}\approx\frac{g\alpha\rho h\Delta T_{\sigma}}{3Ra_{crit}^{1/3}}\end{equation}

Where again $h$ and $\Delta T_{\sigma}$ are the height and driving
temperature contrast of the convective region, respectively. The tensile
stress on the brittle crust may then be amplified if the thickness
of the crust $D$ is smaller than the horizontal length over which
a convection cell is in contact with the crust. Assuming a convection
cell aspect ratio of unity, the stress on the crust is therefore $\tau=\sigma_{conv}h/D$. 

Performing this analysis for a crust in Regime II, we find that even
in the case of small $D$ ($3\, km$) and high $\Delta T_{\sigma}$
($40\, K$), $\tau$ does not rise beyond $220\, kPa$ for $g$ between
10 and $20\, m/s^{2}$. Therefore, the crust in Regime II is probably
not directly broken apart by stresses due to solid-state convection
alone. However, resurfacing may persist in Regime II and even Regime
I crusts due to a range of other phenomena that have been observed
on the large icy satellites of the solar system. Such surface processes
include diapirism (Europa- \citealp{ref59}; Triton- \citealp{ref65}),
cryovolcanism (Titan- \citealp{ref60}), and planetary-scale tectonics
(Ganymede- \citealp{ref66}; Europa- \citealp{ref63}), although the
last of these maybe require exogenic sources of stress such as tidal
forces in addition to internal processes. The presence of solid-state
convection is found to facilitate these forms of resurfacing (e.g.
diapirism- \citealp{ref59}; cryovolcanism-\citealp{ref62}), although
penetration of the icy crust by underlying fluids may still be possible
in the absence of convection \citep{ref68}. Alternatively, a Regime
II crust may be completely devoid of surface activity as in the case
of Callisto \citep{ref63}. 

As considerations of these solar system bodies show, reliably constraining
the outgassing into the atmospheres of Regimes I and II planets is
not possible given only the dynamical and thermal structure of the
ice shell, as presented in this paper. 

On the other hand, in Regimes III and IV, the stress on the lithosphere
is greatly increased for two reasons. First, because the convection
cells extend to the base of the $H_{2}O$ layer of the planet, $\sigma_{conv}$
is much larger due to greater values for $h$ and $\Delta T_{\sigma}$.
Second, assuming convective cells with aspect ratios of near unity,
the amplification from the $h/D$ factor is much greater. We find
that in the $25\%$ and $50\%$ water cases considered in this work,
the crust is easily broken by convection-driven stresses ($\tau\approx10^{3}\, MPa$
for a thick, $50\, km$ crust), leading to continuous global resurfacing
under these Regimes. This result is consistent with analogous analysis
for super-Earths with rocky surfaces; as shown in \citet{ref72},
the higher convective stresses in the mantles of large silicate planets
likewise outcompetes the tensile strength of the brittle crust.

Finally, in the liquid surface case, chemical species released into
the atmosphere should again be selected for water-transportable types,
but outgassing on the surface proceeds unhindered. In this respect,
the properties of outgassed material may be similar to those of Regimes
I and II, while the flux may be similar to that of Regimes III and
IV.

Several uncertainties exist for our model of convection in the ice
mantle. The true viscosity of the Ice VII and Ice X is unconstrained,
as ice viscosity has not been measured at pressures of greater than
about $0.8\, GPa$. The viscosity deep in the ice mantle is determined
by the balance between three processes: increasing pressure, which
increases the activation energy; shrinking activation volume, which
tempers the above effect of pressure; and increasing temperature.
Interestingly, all resulting mantle thermal profiles show a viscosity
maximum in mid-layer of the ice mantle with viscosity values between
$10^{21}$ and $10^{23}\, Pa\cdot s$. This is due to the initial
dominance of the pressure effect over the adiabatic temperature increase
owing to a relatively large activitation volume. However, as the vacancies
of the material become increasingly compressed at depth, the viscosity
decreases with increasing temperature. This effect leads to low BBL
viscosities of as little as $10^{14}\, Pa\cdot s$ for the $10\, M_{Earth}$,
$50\%\, H_{2}O$ planet. Such results may not reflect true values
and improved results must await the availability of additional rheological
data for high-pressure ices. Furthermore, more refined results call
for new theory that characterizes convection with a viscosity maximum
in mid-layer and provides for a means to determine a characteristic
viscosity.

Values for certain other parameters in the mantle Rayleigh number,
such as $g$ and $\alpha$ are difficult to choose to be representative
of the ice mantle. We find that varying these throughout their possible
range generally leads to a change of a factor of a few in the resulting
overturn time. Combined with the greater uncertainty resulting from
the characterization of viscosity as described above, the results
in Table \ref{tab:Overturn-table} are presented as estimates to within
around one order of magnitude instead of precise solutions. We note,
however, that despite these uncertainties, the qualitative phenomenon
of a mid-layer high viscosity region is relatively certain, as the
effect is apparent even when varying the parameters in the viscosity
law beyond the range of published values \citep{ref38}. 

A further point of ambiguity is whether the steep rise in temperature
along the adiabat in the lower parts of the ice mantle will cause
our thermal profile to intersect the melting curve again at depth.
Recent studies of the melting curve of high pressure ice phases suggest
that this may occur only for a planet with a much larger water layer.
\citet{ref50} found experimentally that at $100\, GPa$ corresponding
to a the melting point of the Ice X is 2400 K, while our ice mantle
temperatures are confined to under $1200\, K$ (Figure \ref{fig:Complete-ice-mantle}).

A word must be said about the behavior of material where the temperature
is constrained to the melting curve. This situation arises at the
top of the ice mantle when the crust falls into Regimes I, II, or
IV. In Regimes I and II, a liquid ocean overlies the ice mantle, and
the temperature profile runs along the melt curve until convective
instability is reached in this boundary layer.

Although the globally averaged temperature proceeds along the melt
curve in these cases, the full 3-D scenario is more complicated. Qualitatively,
we expect the upward heat flux in this transition zone to be carried
by the percolation of liquid water in certain concentrated zones.
We describe a feedback mechanism that leads to the creation of narrow
zones of rising meltwater. Beginning with a laterally homogeneous
transition layer, if a packet of material should melt in a specific
location, say due to the presence of an underlying hot plume, this
packet of material immediately undergoes a decrease in density, which
leads to a decrease in the pressure in the underlying column. A decrease
in pressure leads to the melting of ice previously at equilibrium,
further promoting the creation of meltwater in the column. More detailed
considerations of this process should be performed in the future.

\section{Conclusions\label{sec:Conclusions}}

In this work we have modeled the interior dynamics of the $H_{2}O$
layer of water planets. We first study the thermal and dynamical properties
of an ice crust for a planet with sub-freezing $T_{S}$. We find that
the dynamical behavior of the crust falls into one of four Regimes
(Figure \ref{fig:The-four-regimes}), primarily determined by the
planetary parameters surface temperature and heat flux, with a weaker
dependence on surface gravity. Planets with a crust in Regimes III
and IV should show atmospheric signatures of close to the full range
of possible compounds produced from materials released from the silicate
- metal core. Ice tectonics under these Regimes is expected to bring
about continual resurfacing. The atmosphere of Regimes I and II planets
may or may not be enriched by outgassing, depending on the presence
of resurfacing phenomena such as cryovolcanism and other sources of
stress (e.g. tides). As evidenced by the examples of such bodies studied
in our solar system, the surface geology of icy bodies in these Regimes
is difficult to constrain without case-specific considerations of
a wider range of resurfacing processes.

In the case of Regimes I and II, the range of materials that is able
to reach the surface is affected by the presence of a liquid ocean
layer, which selects for materials that can exist in solution or suspension.
A more detailed study of this process is important for work future
work. In the case of a liquid surface layer, we expect a full flux
of outgassed materials biased by its transportability through the
liquid ocean.

Our model of the ice mantle composed of higher-phase ices shows that
materials expelled from the silicate - metal core into the ice layer
can traverse the ice mantle on the time scale of $0.1$ to a hundred
$Myr$ for the cases examined with shorter overturn times corresponding
to larger planets. New constraints on the rheology of higher-phase
ices as well as theory to describe the behavior of convection with
a mid-layer viscosity maximum are required to obtain more accurate
results. The specific chemical composition and mass flux of such dredged
and outgassed material should be investigated quantitatively in further
studies, ultimately leading to specific predictions of the atmospheric
detectables of water planets.

\section{Acknowledgements}

We thank Diana Valencia, Lisa Kaltenegger and Wade Hennings for discussions
that led to advances in understanding and the conception of the scope
of this paper. This work is supported by the Harvard University Origins
of Life Initiative. 

\clearpage

\begin{table}[H]
\begin{centering}
\begin{tabular}{ccccccc}
\hline 
Component & EOS & $\rho_{0}\,(\frac{kg}{m^{3}})$ & $K_{0T}\,(GPa)$ & $K'_{0T}$ & $K''_{0T}\,(GPa^{-1})$ & Ref.\tabularnewline
\hline
$Fe$ & Vinet & 8300 & 156.2 & 6.08 & - & 1\tabularnewline
$MgSiO_{3}$ & B-M 4 & 4100 & 247 & 3.97 & -0.016 & 2\tabularnewline
$Ice\, X$ & B-M 3 & 1239 & 4.26 & 7.75 & - & 3\tabularnewline
$Ice\, VII$ & B-M 3 & 1463 & 23.7 & 4.15 & - & 4\tabularnewline
$Ice\, Ih$ & B-M 3 & 917 & 9.86 & 6.6 & - & 5, 6\tabularnewline
\hline
\end{tabular}
\par\end{centering}

\caption{Equations of state and elastic parameters used in calculation of static
planetary dimensions. B-M 3 and B-M 4 refers to the 3rd and 4th order
Birch-Murnaghan equations of state while $\rho_{0}$ and $K_{0T}^{n}$
are the zero-pressure density and pressure derivatives of the isothermal
bulk modulus, repectively. \label{tab:parameters for statics}}

References: (1) \citealp{ref19} (2) \citealp{ref20} (3) \citealp{ref21}
(4) \citealp{ref22} (5) \citealp{ref23} (6) \citealp{ref24}
\end{table}

\begin{table}[H]
\begin{centering}
\begin{tabular}{cccc}
\hline 
Constant & $0.02\%\, H_{2}O$ & $25\%\, H_{2}O$ & $50\%\, H_{2}O$\tabularnewline
\hline
$A$ & $0.994$ & $1.157$ & $1.255$\tabularnewline
$B$ & $0.266$ & $0.253$ & $0.251$\tabularnewline
\hline
\end{tabular}
\par\end{centering}

\caption{Constants for the mass-radius scaling relation (Equation \ref{eq:scaling mr}).
\label{tab:Constants-for-the}}

\end{table}

\begin{table}[H]
\begin{centering}
\begin{tabular}{cccc}
\hline 
$H_{2}O$ Content & $2\, M_{Earth}$ & $5\, M_{Earth}$ & $10\, M_{Earth}$\tabularnewline
\hline
$25\%\, H_{2}O$ & $47.8\pm1.0\, GPa$ & $115.3\pm1.6\, GPa$ & $236.6\pm2.6\, GPa$\tabularnewline
$50\%\, H_{2}O$ & $86.2\pm2.0\, GPa$ & $212.0\pm2.2\, GPa$ & $436.9\pm3.6\, GPa$\tabularnewline
\hline
\end{tabular}
\par\end{centering}

\caption{Pressures at the ice - silicate boundary. Values greater than about
$60\, GPa$ indicate the prescence of Ice X. Errors are determined
by using the extremes of each equation of state parameter that yields
the lowest and highest pressure. \label{tab:Pressures-at-the}}

\end{table}

\begin{table}[H]
\begin{centering}
\begin{tabular}{cccccc}
\hline 
Transition & $\Delta\rho\,(kg/m^{3})$  & $\gamma\,(MPa\, K^{-1})$ & $P_{b}$ & Ref. for $\Delta\rho$ & Ref. for $\gamma$\tabularnewline
\hline
$Ih-III$ & $218$ & $-0.47$ - $-0.24$ & $-0.01$ - $-5\times10^{-3}$ & 1 & 2; 3\tabularnewline
$II-V$ & $38$ & $-8.9$ - $-0.67$  & $-0.025$ - $-.002$ & 1, 4 & 2; 3, 5\tabularnewline
$II-VI$ & $48$ & $\approx-1.7$ & $\approx-5\times10^{-3}$ & 1, 4 & 6\tabularnewline
$VII-X$ & $68.1$ & $-11$ - $+26.3$ & $-0.01$ - $+0.02$ &  7, 8 & 9; 10\tabularnewline
\hline
\end{tabular}
\par\end{centering}

\caption{The phase buoyancy parameter $P_{b}$ for all encountered endothermic
phase transitions between the water ices. Throughgoing convection
is unhindered if $P_{b}\gtrsim-0.3$. Since the Clapeyron slope $\gamma$
is the most uncertain parameter in $P_{b}$ (see Equation \ref{eq:P}),
the range of values of $P_{b}$ is found by evaluation with a range
of $\gamma$ values found in the literature. \label{tab:The-buoyancy-parameter}}

References: (1) \citealp{ref44} (2) \citealp{ref35} (3) \citealp{ref42}
(4) \citealp{ref45} (5) \citealp{ref43} (6) \citealp{ref38} (7)
\citealp{ref21} (8) \citealp{ref22} (9) \citealp{ref28} (10) \citealp{ref30}
\end{table}

\begin{table}[H]
\begin{centering}
\begin{tabular}{cccc}
\hline 
$H_{2}O$ Content & $2\, M_{Earth}$ & $5\, M_{Earth}$ & $10\, M_{Earth}$\tabularnewline
\hline
$25\%\, H_{2}O$ & $83\, Myr$ & $16\, Myr$ & $0.8\, Myr$\tabularnewline
$50\%\, H_{2}O$ & $190\, Myr$ & $12\, Myr$ & $0.3\, Myr$\tabularnewline
\hline
\end{tabular}
\par\end{centering}

\caption{Order of magnitude estimates for the overturn time scale for the ice
mantles. Numbers represent the time for material to rise from ice
- silicate boundary to the top of the ice mantle. The values for the
$10\, M_{Earth}$ planets are not as well-constrained. See the Discussion
section for further notes about uncertainties.\label{tab:Overturn-table}}

\end{table}


\begin{thebibliography}{CRC Handbook (1969}
\bibitem[Anderson et al (2001)]{ref19}Anderson, O. L., Dubrovinsky,
L., Saxena, S. K., \& LeBihan, T. 2001, Geophys. Res. Lett., 28, 399

\bibitem[Barr and McKinnon (2007)]{ref58}Barr, A. C. and McKinnon,
W. B. 2007, J. Geophys. Res, 112, E02012

\bibitem[Bridgman (1912)]{ref42}Bridgman, P. W. 1912, Proc. Am. Acad.
Arts Sci., 47, 439

\bibitem[Caracas et al. (2008)]{ref52}Caracas, R., \& Cohen, R. E.
2008, Phys. Earth Planet. Inter., 168, 147

\bibitem[Christensen and Yuen (1985)]{ref41}Christensen, U. R., \&
Yuen, D. A. 1985, JGR, 90, 10291

\bibitem[Dumoulin et al. (1999)]{ref57}Dumoulin, C. et al. 1999,
J. Geophys. Res., 104, B6, 12759

\bibitem[Durham et al. (1997)]{ref38}Durham, W. B., Kirby, S. H.,
\& Stern, L. A. 1997, J. Geophys. Res., 102, 16293

\bibitem[Durham and Stern (2001)]{ref73}Durham, W. B., Stern, L.
A., 2001, Annu. Rev. Earth Planet. Sci., 29, 295

\bibitem[Ehrenreich and Cassan (2007)]{ref15}Ehrenreich, D., \& Cassan,
A. 2007, Astron. Nachr., 328, 789

\bibitem[Fei et al. (1993)]{ref40}Fei, Y., Mao, H., \& Hemley, R.
J. J. Chem. Phys., 99, 5369

\bibitem[Fortney et al. (2007)]{ref12}Fortney, J., Marley, M., \&
Barnes, J. 2007, ApJ, 659, 1661

\bibitem[Goncharov et al. (1999)]{ref30}Goncharov, A. F., Struzhkin,
V. V., Mao, H., \& Hemley, R. J. 1999, Phys. Rev. Letters, 83, 1998

\bibitem[Grasset et al. (2009)]{ref10}Grasset, O., Schneider, J.,
Sotin, C. 2009, ApJ, 693, 722

\bibitem[Head et al (2002)]{ref66}Head, J. et al. 2002, Geophys.
Res. Lett. 29, 24, 2151

\bibitem[Hemley et al (1987)]{ref22}Hemley, R. J. et al. 1987, Nature,
330, 737

\bibitem[Howard (1966)]{ref74}Howard, L. N. 1966, in Proceedings
of the 11th Congress of Applied Mechanics, Munich (Germany), ed Gortler,
H., 1966 (Berlin: Springer-Verlag), 1109-1115

\bibitem[Ida and Lin (2004)]{ref3}Ida, S., \& Lin, D. N. C. 2004,
ApJ, 604, 388

\bibitem[Karki and Wentzcovitch (2000)]{ref20}Karki, B. B., Wentzcovitch,
R. M., de Gironcoli, S., \& Baroni, S. 2000, Phys Rev. B, 62, 14750

\bibitem[Kennedy et al. (2006)]{ref5}Kennedy, G., Kenyon, S., \&
Bromley, B. 2006, ApJ, 650, L139

\bibitem[Kuchner (2003)]{ref6}Kuchner, M. 2003, ApJ, 596, L105

\bibitem[Leger et al. (2004)]{ref7}Leger, A., et al. 2004, Icarus,
169, 499

\bibitem[Leger et al. (2009)]{ref2}Leger, A., et al. 2009, arXiv:0908.0241

\bibitem[Loubeyre et al. (1999)]{ref21}Loubeyre, P., LeToullec, R.,
Wolanin, E., Hanfland, M., \& Hausermann, D. 1999, Nature, 397, 503

\bibitem[Hogenboom (1997)]{ref70}Hogenboom, D. L. et al. 1997, Icarus,
128, 171

\bibitem[Leon et al. (2002)]{ref45}Cruz Leon, G., Rodriguez Romo,
S., \& Tchijov, V. 2002, J. Phys. and Chem. of Solids, 63, 843

\bibitem[Lobban et al. (1998)]{ref29}Lobban, C., Finney, J. L., \&,
Kuhs, W. F. 1998, Nature, 391, 268

\bibitem[Lopes et al (2007)]{ref60}Lopes, R. M. C. et al. 2007, Icarus,
186, 395

\bibitem[Mashimo et al. (2006)]{ref25}Mashimo, T. et al. 2006, Phys.
Rev. Letters, 96, 105504

\bibitem[Mayor et al. (2009)]{ref51}Mayor, M. et al. 2009, arXiv:0906.2780

\bibitem[McKinnon (1998)]{ref35}McKinnon, W. B. 1998, in Solar System
Ices, ed Schmitt, B., de Bergh, C., \& Festou, M. 1998, (Dordrecht:
Kluwer Academic Publishers), 525

\bibitem[McKinnon (2006)]{ref53}McKinnon, W. B. 2006, Icarus, 183,
435

\bibitem[Melnick et al. (2001)]{ref1}Melnick et al. 2001, 199th AAS
Meeting, \#09.10, in Bulletin of the American Astronomical Society,
34, 559

\bibitem[Mercury et al. (2001)]{ref43}Mercury, L., Vieillard, P.,
\& Tardy, Y. 2001, Appl. Geochem., 16, 161

\bibitem[Mitri et al. (2008)]{ref62}Mitri, G. et al. 2008, Icarus,
196, 216

\bibitem[Mueller and McKinnon (1988)]{ref33}Mueller, S., \& McKinnon,
W. B. 1988, Icarus, 76, 437

\bibitem[Nimmo and Schenk (2006)]{ref47}Nimmo, F., \& Schenk, P.
2006, J. Struc. Geol., 28, 2194

\bibitem[O'Connell (1977)]{ref39}O'Connell, R. J. 1977, Tectonophysics,
38, 119

\bibitem[O'Connell and Hager (1980)]{ref75}O'Connell, R. J. and Hager,
B. H. 1980, in Proceedings of the International School of Physics
'Enrico Fermi'; course 78, ed Dziewonski, A. M. and Boschi, E., 1980,
(North Holland), 270

\bibitem[Pappalardo et al (1998)]{ref59}Pappalardo, R. T. et al.
1998, Nature, 391, 365

\bibitem[Raymond et al. (2004)]{ref4}Raymond, S. N., et al. 2004,
Icarus, 168, 1

\bibitem[Schubert et al. (2001)]{ref36}Schubert, G., Turcotte, D.
L., \& Olson, P. 2001, Mantle Convection in the Earth Planets, (Cambridge:
Cambridge Univ. Press), 422

\bibitem[Schwager et al. (2004)]{ref50}Schwager, B., Chudinovshkikh,
L., Gavriliuk, A., \& Boehler, R. 2004, J. Phys. Condens. Matter,
16, S1177

\bibitem[Seager et al. (2007)]{ref13}Seager, S., Kuchner, M., Hier-Majumder,
C. A., \& Militzer, B. 2007, ApJ, 669, 1279

\bibitem[Selsis et al. (2007)]{ref14}Selsis, F., et al. 2007, Icarus,
191, 453

\bibitem[Shankar et al. (1999)]{ref18}Shankar J., Kushwah, S. S.,
\& Sharma, M. P. 1999, Physica B, 271, 158

\bibitem[Shaw (1986)]{ref44}Shaw, G. H. 1986, J. Chem. Phys., 84,
5862

\bibitem[Schenk and Jackson (1993)]{ref65}Schenk, P. and Jackson,
M. P. A. 1993, Geology, 21, 299

\bibitem[Showman and Malhotra (1999)]{ref63}Showman, A. P. and Malhotra,
R. 1999, Science, 286, 77

\bibitem[Showman et al (2004)]{ref68}Showman, A. P. et al. 2004,
Icarus, 172, 625

\bibitem[Solomatov (1995)]{ref54}Solomatov, V. S. 1995, Phys. Fluids,
7(2), 266

\bibitem[Solomatov and Moresi (2000)]{ref56}Solomatov, V. S. and
Moresi, L. -N. 2000, J. Geophys. Res. 105, B9, 21795

\bibitem[Song et al. (2003)]{ref28}Song, M. et al. 2003, Phys. Rev.
B, 68, 014106

\bibitem[Sotin et al. (2007)]{ref9}Sotin, C., Grasset, O., Mocquet,
A. 2007, Icarus, 191, 337

\bibitem[Spohn and Schubert (2003)]{ref31}Spohn, T. \& Schubert,
G. 2003, Icarus, 161, 456

\bibitem[Strassle et al. (2005)]{ref24}Strassle, Th., Klotz, S.,
Loveday, J. S., \& Braden, M. 2005, J. Phys. Condens. Matter, 17,
3029

\bibitem[Tajika (2008)]{ref16}Tajika, E. 2008, ApJ, 680, L53

\bibitem[Turcotte and Schubert (2002)]{ref71}Turcotte, D. L., \&
Schubert, G. 2002, Geodynamics, (New York: Cambridge University Press),
136

\bibitem[Unemoto et al (2008)]{ref27}Unemoto, K., Wentzcovitch, R.
M., \& Allen, P. B. 2008, Science, 311, 983

\bibitem[Valencia et al. (2007a)]{ref8}Valencia, D., Sasselov, D.,
\& O'Connell, R. 2007, ApJ, 656, 545

\bibitem[Valencia et al. (2007b)]{ref11}Valencia, D., Sasselov, D.,
\& O'Connell, R. 2007, ApJ, 665, 1413

\bibitem[Valencia et al. (2007c)]{ref46}Valencia, D., O'Connell,
R. J., \& Sasselov, D. 2007, ApJ, 670, L45

\bibitem[Valencia and O'Connell (2009)]{ref72}Valencia, D., O'Connell,
R. G., 2009, Earth Planet. Sci. Lett. 286, 492 

\bibitem[Vinet and Ferrante (1987)]{ref17}Vinet, P., \& Ferrante,
J. 1987, J. Geophys. Res., 92, 9319a

\bibitem[CRC Handbook (1969]{ref23}Weast, R. C. 1969, in The CRC
Handbook of Chemistry and Physics, (Cleveland: The Chemical Rubber
Company), F1 

\end{thebibliography}
\end{document}